# Low-Loss Superconducting Resonators Fabricated from Tantalum Films Grown at Room Temperature

Guillaume Marcaud,[1, *] David Perello,[1] Cliff Chen,[1] Esha Umbarkar,[1] Conan Weiland,[2] Jiansong Gao,[1] Sandra Diez,[1, †] Victor Ly,[1] Neha Mahuli,[1] Nathan D'Souza,[1] Yuan He,[1] Shahriar Aghaeimeibodi,[1] Rachel Resnick,[1, ‡] Cherno Jaye,[2] Abdul K. Rumaiz,[3] Daniel A. Fischer,[2] Matthew Hunt,[1] Oskar Painter,[1] and Ignace Jarrige[1, §]

[1]*AWS Center for Quantum Computing, Pasadena, CA 91106, USA*
[2]*Materials Measurement Science Division, Material Measurement Laboratory, National Institute of Standards and Technology, Gaithersburg, MD 20899, USA*
[3]*National Synchrotron Light Source II, Brookhaven National Laboratory, Upton, NY 11973*

(Dated: January 16, 2025)

The use of $\alpha$-tantalum in superconducting circuits has enabled a considerable improvement of the coherence time of transmon qubits. The standard approach to grow $\alpha$-tantalum thin films on silicon involves heating the substrate, which takes several hours per deposition and prevents the integration of this material with wafers containing temperature-sensitive components. We report a detailed experimental study of an alternative growth method of $\alpha$-tantalum on silicon, which is achieved at room temperature through the use of a niobium seed layer. Despite a substantially higher density of oxygen-rich grain boundaries in the films sputtered at room temperature, resonators made from these films are found to have state-of-the-art quality factors, comparable to resonators fabricated from tantalum grown at high temperature. This finding challenges previous assumptions about correlations between material properties and microwave loss of superconducting thin films, and opens a new avenue for the integration of tantalum into fabrication flows with limited thermal budget.

## I. INTRODUCTION

Superconducting qubit technology is now an established front runner towards the realization of a universal quantum computer. However, scalability remains a central challenge for superconducting quantum circuits as thousands of physical qubits are required to generate a logical qubit through quantum error correction. In order to reduce the scaling overhead, it is crucial to lower the error rates of the individual physical qubits. Research towards this goal has shown that improvements in the quality of the material surfaces and interfaces, which intrinsically host coherence-limiting defects, are linked to longer qubit coherence times.

The most notable breakthrough in this area of research came from the adoption of the high-critical-temperature phase of tantalum, $\alpha$-Ta (T$_c \sim$ 4.4 K) [1], as the base superconducting layer in transmon qubits [2–4]. The lower microwave loss of $\alpha$-Ta compared with other elemental superconductors traditionally used in the fabrication of transmon qubits, such as Al and Nb, has been ascribed to the thin and stable native oxide layer that is primarily made of $Ta_2O_5$ [5]. There are two sputtering-based methods known to form $\alpha$-Ta. One involves deposition at temperatures exceeding 350°C, and the other at room temperature with the use of a seed layer such as Nb [6, 7]. Published studies that rely on sputtering to fabricate Ta-based resonators and qubits suggest that only devices made from films deposited at high temperature (HT) benefit from the characteristic performance boost. Indeed, reported internal quality factors for resonators made from Ta films sputtered at room temperature (RT) on sapphire [8] and silicon [9–12] are in the $10^5$ range in the single photon regime, which is about one order of magnitude lower than devices made from HT Ta films [13–16]. However, the HT deposition process has its own drawbacks, such as the incompatibility with low-thermal-budget fabrication flows, the low throughput due to the heating and cooling time, and the potential for additional loss channels associated with the metal-substrate interdiffusion or contamination from outgassing in the deposition chamber.

Here, we report on the fabrication and measurement of Ta-based coplanar waveguide (CPW) resonators made from films sputtered at RT and HT on silicon (001) substrate. The resonators share the same design, measurement conditions, and fabrication process besides the sputtering step, which ensures a clean comparison between the microwave loss properties of the films. We show that the TLS loss is comparable for resonators made from both types of films despite clear differences in their microstructural, chemical, and transport properties. The implications of this finding are twofold. First, it shows that the RT Ta sputtering process is viable for the fabrication of state-of-the-art superconducting circuits. Second, the absence of clear correlation between the film properties and resonator performance highlights the robustness of Ta surface loss tangents against variations in the deposition process.

---

[*] Corresponding email address: gumarcau@amazon.com
[†] Current affiliation: National Institute of Standards and Technology
[‡] Current affiliation: Google Research
[§] Corresponding email address: jarrige@amazon.com

## II. RESULTS

### A. Resonator design and measurements

In order to assess the influence of the sputtering method on the performance of superconducting circuits patterned into the Ta films, $\lambda/4$ CPW resonators in the 4.5-6.5 GHz frequency range were fabricated from each type of film, RT and HT, and two design splits were chosen with the gap and trace width ($w_g = w_{tr} = w$) set to $w = 3$ $\mu$m and $w = 30$ $\mu$m. The two geometries provide a different set of participation ratios for the relevant dielectric regions (substrate, air) and interfaces (metal-air, metal-substrate, substrate-air). See Supplementary Material for further details on the design of the resonators and simulation of the participation ratios. This allows us to finely compare the microwave loss between the two types of Ta films when the interface participation ratio is increased for $w = 3$ $\mu$m, and assess whether other loss mechanisms start to limit the quality factors of the resonators when the interface participation ratio is reduced for $w = 30$ $\mu$m.

Figure 1 presents the loss measured at high power ($\delta_{hp} = 1/Q_{hp}$), at low power in the single-photon regime ($\delta_{1ph} = 1/Q_{1ph}$), and the inferred two-level system (TLS) induced loss ($\delta_{tls} = 1/Q_{tls}$) for the four material-design configurations. The TLS loss is usually described with the standard tunneling model (STM) [17–19] that is characterized by a dependence on electric field intensity, $\delta_{tls} \sim (1 + E^2/E_c^2)^{-1/2}$, where $E$ is the electric field and $E_c$ denotes the critical field at which the TLS bath saturates. As the electric field intensity seen by the average TLS can be related to the circulating power within the resonator, this model allows us to extract $\delta_{tls}$ from the difference between the loss at low and high power, as $\delta_{tls} = \delta_{1ph} - \delta_{hp}$. The accuracy of this method is contingent upon saturating the TLS response in both power regimes, a condition which is not consistently met in our data, especially for $w = 3$ $\mu$m. Accordingly, we limit our use of this metric to a coarse comparison of $\delta_{tls}$ between the two materials.

Despite identical designs, the resonators fabricated from HT and RT Ta films have different resonance frequencies, which is particularly evident for $w = 3$ $\mu$m in Figure 1 (c) and (d). This shift is induced by the kinetic inductance ($L_k$) of the films that we estimate to be approximately 400 fH/sq for RT Ta and 50 fh/sq for HT Ta. The medians, interquartile ranges (iqr), and standard deviations (std) of $\delta_{1ph}$ and $\delta_{tls}$ are calculated from all the measurements of all resonators for each material and design and summarized in Table I. The losses, which are measured between 15 and 70 times per resonator, are plotted as a function of the resonance frequencies to highlight the spread caused by the temporal variations, and rule out any frequency dependence.

We find that $\delta_{1ph}$ for HT and RT Ta films match state-of-the-art performance reported for similar designs and frequency on silicon [13], with median values of $\delta_{1ph}$ on the order of $2 \times 10^{-7}$ for $w = 30$ $\mu$m and $3.5 \times 10^{-7}$ for $w = 3$ $\mu$m. A more detailed comparison with the literature is presented in Supplementary Material. A statistically significant increase in $\delta_{tls}$ is observed from $w = 30$ $\mu$m to $w = 3$ $\mu$m for both materials due to the larger interface participation ratio for narrower trace and gap. However, the median $\delta_{tls}$ remain comparable between the two materials for each width, with differences of about 20% at $w = 30$ $\mu$m and 50% at $w = 3$ $\mu$m, which is on par with the interquartile range and total standard deviation of the measurement for each material and design.

TABLE I. Median, interquartile range (iqr), and standard deviation (std) of $\delta_{1ph}$ and $\delta_{tls}$ for the two materials, HT and RT, and the two geometries, $w = 30$ $\mu$m and $w = 3$ $\mu$m.

| | $\delta_{1ph}$ ($\times 10^{-7}$) | | | $\delta_{tls}$ ($\times 10^{-7}$) | | |
| --- | --- | --- | --- | --- | --- | --- |
| | median | iqr | std | median | iqr | std |
| HT - 30 $\mu$m | 2.4 | 1.7 | 1.4 | 1.2 | 0.4 | 0.4 |
| RT - 30 $\mu$m | 1.6 | 1.2 | 0.9 | 1.0 | 0.4 | 0.2 |
| HT - 3 $\mu$m | 2.9 | 1.3 | 1.0 | 2.5 | 1.2 | 1.0 |
| RT - 3 $\mu$m | 4.1 | 1.7 | 1.2 | 3.7 | 1.7 | 1.2 |

Our finding that both materials have comparable TLS loss is at odds with two recent reports of higher losses in resonators fabricated from RT Ta compared with HT Ta. In the first report [8], losses were found to be three times larger for RT Ta on a Nb underlayer compared with HT Ta, which was tentatively assigned to the smaller grains of the RT film. Potentially contributing to the diverging results is that our chips were treated in a buffered oxide etchant (BOE) prior to their packaging and measurement, while those in Ref. [8] were not. A scenario where a pre-packaging BOE treatment would be required to achieve comparable performance between these two materials could suggest that the surface of the RT Ta films is more prone to fabrication-induced defects, which results in higher losses unless the defect-rich surface layers are removed by BOE. The second report [9] showed that $Q_{1ph}$ for RT Ta films grown on TiN and TaN underlayers is respectively two and seven times smaller than HT Ta. Here, a BOE treatment was applied to all devices prior to packaging. This finding suggests that the material used for the seed layer directly contributes to the dielectric loss of the devices, and that TiN and TaN grown at room temperature can be more lossy than Nb, although we cannot exclude that a careful optimization of the growth parameters of the nitride seed layers could lead to a reduction of their TLS loss [20].

### B. Materials characterization: structure, morphology, and transport properties

The comparable loss between HT and RT allows us to test long-standing hypotheses regarding the influence of the film microstructure, transport properties, and surface and interface chemistry on the microwave loss [21]. In this section, we focus on the first two. We start with the bulk structural properties of the films measured by x-ray diffraction (XRD). The XRD patterns, shown in Figure 2 (a), confirm that there is no trace of spurious $\beta$-Ta phase in either film. One of the main difference between the two films is the three-times larger full width at half maximum ($\Delta$) of the (hh0) specular reflections along the $2\theta$ axis for RT Ta compared with HT Ta. This can be the result of smaller coherent domain sizes along the growth direction or stronger strain gradient in RT Ta. Another difference between the two films is the presence of (310) and (321) parasitic out-of-plane orientations in the HT film besides (110), whereas (110) is the only orientation detected in the RT film. We also note the presence of rings in the pole figures in Figure 2 (b) and (c) that demonstrate the lack of long-range in-plane ordering in either film.

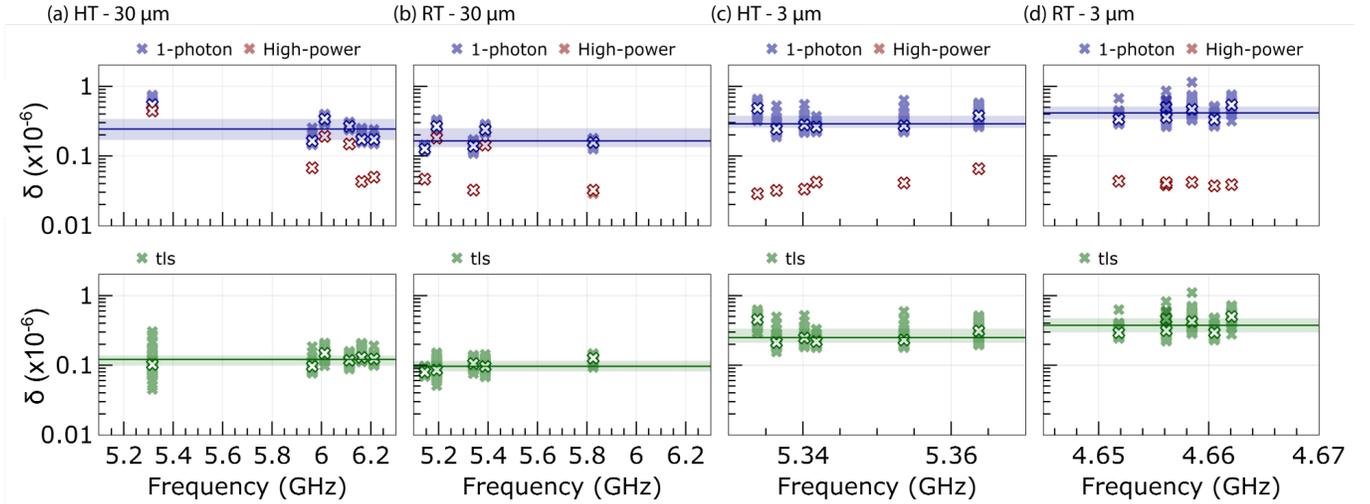

FIG. 1. Loss measurements at high power ($\delta_{hp}$) and low power in the single-photon regime ($\delta_{1ph}$) (top), and calculated loss induced by two-level systems ($\delta_{tls} = \delta_{lp} - \delta_{hp}$) (bottom) as function of resonant frequency for (a) HT Ta - $w = 30$ μm, (b) RT Ta - $w = 30$ μm, (c) HT Ta - $w = 3$ μm, and (d) RT Ta - $w = 3$ μm. The medians $\delta_{1ph}$ and $\delta_{tls}$ of each resonator are indicated by white markers and the median and interquartile range, also reported Table I, are represented by a horizontal line and a strip, respectively.

The distinct surface morphology of the films are observed using atomic-force microscopy (AFM) Figure 3 (a) and (b). The surface of HT Ta is characterized by micron-size pyramidal structures with a highest peak-to-valley height ($S_z$) of 11 nm, whereas the RT Ta surface shows worm-like features with an $S_z$ of 24 nm. Despite similar root-mean-square roughness, 1.28 nm for HT Ta and 1.32 nm for RT Ta, a larger relative increase of surface area is found for RT Ta (3% for RT and <0.01% for HT) due to the larger $S_z$ and smaller feature size, which, in turn, results in a larger effective dielectric region at the surface of the RT Ta film that can contribute to the surface loss tangent.

Complementary information regarding the microstructure and morphology of the films can be obtained using electron backscattering diffraction (EBSD) and transmission Kikuchi diffraction (TKD) techniques. EBSD data for HT Ta, shown in Figure 3 (c) and (d), reveals a variable orientation grain-to-grain but also within grains, and a median grain size of 15 μm that is estimated from the false-color, larger-scale grain map shown in (d). TKD was used for the smaller-grain RT film due to its higher spatial resolution than EBSD. A map of the growth direction is shown in (e) from a cross-section view of a lamella using the same color code as in (c). The grains are confirmed to grow in column along the [hh0] direction as represented in green, and the lateral grain size is found to be about three orders of magnitude smaller than HT Ta, approximately 15 nm.

The structure and morphology of the films also translate into the transport properties presented in Table II. While $T_c$ remains similar, the residual resistance ratio ($RRR$) of RT Ta is found to be an order of magnitude smaller than HT Ta as expected from the higher density of grain boundaries in the RT Ta film. The sheet kinetic inductance ($L_k$), which inversely depends on the superfluid density, is calculated from the sheet resistance measured above the critical temperature at 4.5 K ($R_{4.5K}$) on the resonators (>3000 squares) with $L_k = \hbar R_{5K}/1.76\pi k_b T_c$ [22]. The difference in $L_k$ by more than an order of magnitude between the two films match the shift in resonance frequency discussed in the previous section and suggests a lower superfluid density in RT Ta, in agreement with the idea of a higher density of defects and shorter mean free path in this film [23]. A larger $L_k$ also means a greater kinetic inductance fraction and therefore a pronounced sensitivity of the resonator frequency to material non-uniformity such as resistivity or thickness. Other transport properties such as the carrier density and mean free path in the normal state, penetration depth, critical field and superconducting coherence length are given in the Supplementary Material for future reference.

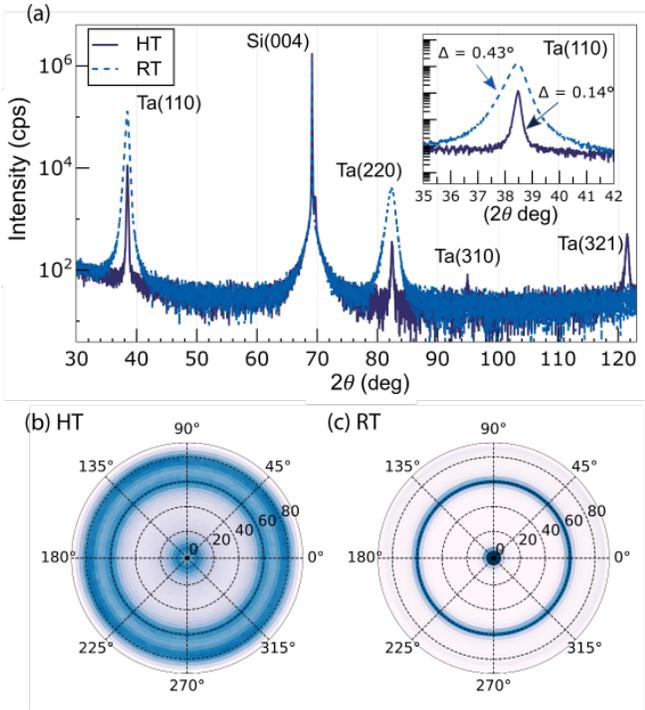

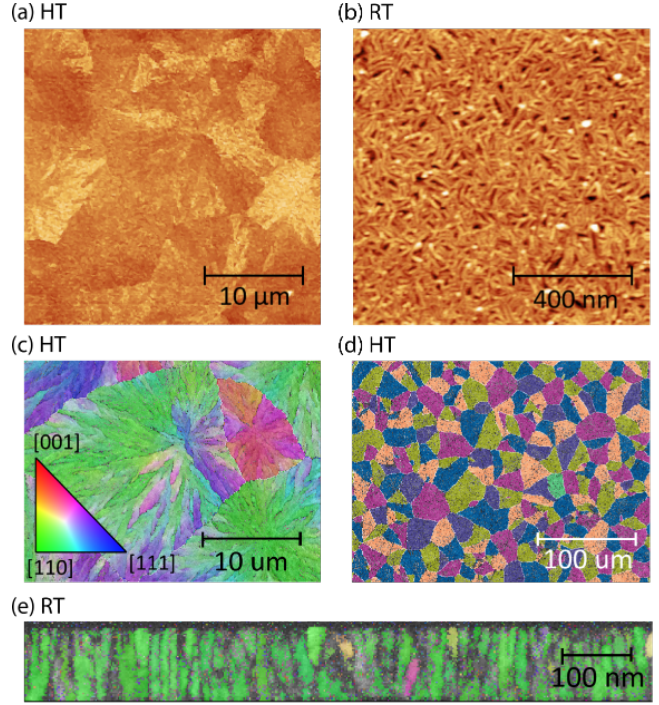

FIG. 2. (a) X-ray diffraction 2θ-θ scans showing the growth direction of the tantalum for the two films. A zoomed-in view of the Ta(110) specular reflection is shown in inset. (b), (c) X-ray diffraction pole figures for HT and RT Ta films, respectively. The color-scale represents the intensity of the Ta(110) reflection, with a maximum in dark blue. The ring centered around $\omega=75°$ is assigned to a parasitic in-plane reflection.

FIG. 3. (a), (b), AFM topography images of HT and RT Ta films surface, with different field of view and color scale. (c), (d) Electron back-scattering diffraction maps showing (c) the color-coded growth direction of few HT Ta grains and (d) the false-color large scale view of HT Ta grains. (e) Transmission Kikuchi diffraction map of a cross section of a RT Ta film. The color code represents the growth direction as in (c).

TABLE II. Transport properties of HT and RT Ta films. Critical temperature ($T_c$), sheet resistance at 300 K and 4.5 K ($R_{300K}$, $R_{4.5K}$), residual resistance ratio ($RRR$) and sheet kinetic inductance ($L_k$).

|       | $T_c$ (K) | $R_{300K}$ ($\Omega/\square$) | $R_{4.5K}$ ($\Omega/\square$) | $RRR$ | $L_k$ ($fH/\square$) |
|-------|-----------|-------------------------------|-------------------------------|-------|----------------------|
| HT Ta | 4.30      | 1.57                          | 0.078                         | 20.4  | 25                   |
| RT Ta | 4.45      | 2.94                          | 1.16                          | 2.6   | 360                  |

### C. Materials characterization: surface and interface chemistry

We now turn to the analysis of the native oxide layers and the metal-substrate interfaces. We start by a baseline characterization of the oxide species present at the surface of HT and RT Ta films treated with BOE using variable kinetic energy X-ray photoelectron spectroscopy (VKE-XPS). The relative concentration of the different oxidation states of Ta as a function of depth, shown in Figures 4 (a) and (b), is derived from the deconvolution of a set of Ta $4f$ XPS data measured for 14 values of the incident photon energy between 400 eV and 6000 eV (see the Supplementary Material for a detailed description of the data analysis). For both films, the native oxide is found to have an outermost $Ta_2O_5$ layer and a transition layer of suboxides at the interface with Ta, in agreement with recent studies of the surface chemistry of Ta [5, 24]. The RT Ta film shows a thicker native oxide overall and a thicker suboxide layer in particular, which is confirmed by the values of the effective oxide thickness ($h_{ox}$) calculated from a multi-layer model also shown in Figures 4 (a) and (b).

Insights into the spatial distribution of the native oxide regions in fabricated devices are provided by the oxygen EELS maps in Figures 4 (c) and (d) and the corresponding line scans shown in the Supplementary Material. The measurements were carried out on cross-sections of resonator traces made from HT Ta and RT Ta films. Fabrication appears to have caused the native oxide to grow thicker for both films, albeit in substantially different ways. While both Ta traces present a 5-nm thick O-rich outermost layer, the interface between this layer and Ta underneath is sharp for HT Ta but presents Ta-O intermixing in RT Ta, primarily along the grain boundaries. We note that suboxides had also been shown to form along the grain boundaries of Nb films [25]. Turning to the metal-substrate interface, a 5-nm thick amorphous interdiffusion layer between Ta and Si is observed in the HT Ta resonator, in contrast with the sharper Ta/Nb/Si interfaces of the RT Ta resonator. This interdiffusion layer is a likely host for disordered Ta-Si alloys and struc-tural defects [26, 27].

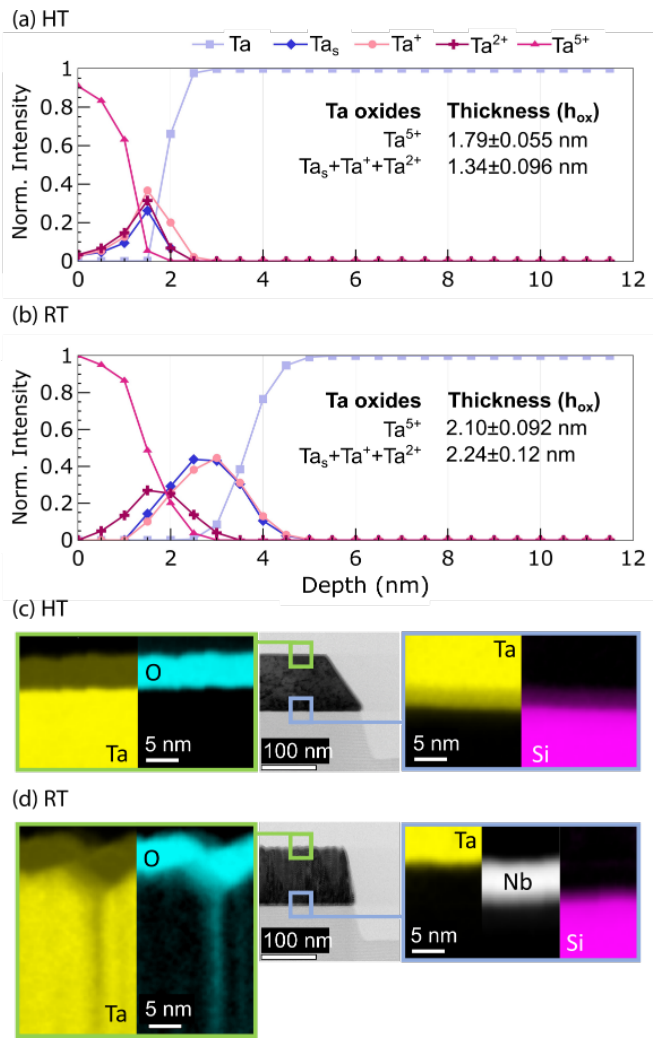

FIG. 4. (a), (b) Depth profile of the tantalum oxide composition at the surface of HT Ta (a) and RT Ta (b) films, extracted from energy-dependent x-ray photoelectron spectroscopy. Estimates of the oxide layer thicknesses ($h_{ox}$) and their error bars, associated to different Ta valence states and derived from a multi-layer model, are also shown in (a) and (b). (c), (d) Bright-field transmission electron microscopy view (TEM) and electron energy-loss spectroscopy (EELS) composition analysis of the cross-section of a resonator made from HT Ta (c) and RT Ta (d).

## III. DISCUSSION

Two clear differences between the properties of the RT Ta and HT Ta films emerge from our multimodal characterization study. On the one hand, the higher density of grain boundaries in the RT Ta films, caused by an average grain size smaller by three orders of magnitude than HT Ta, creates pathways for oxygen diffusion. In devices fabricated from RT Ta films, this results in oxygen contamination and the formation of amorphous Ta suboxides extending several tens of nm below the surface along the grain boundaries. On the other hand, the RT Ta films benefit from a sharper metal-substrate interface than the HT films, which show a Ta-Si interlayer. Both oxygen diffusion in RT Ta and Ta-Si interdiffusion in HT Ta result in an increase of the density of local chemical and structural defects that are commonly associated with dielectric loss.

Although we cannot exclude the possibility of having comparable contributions to loss from the above-mentioned phenomena, effectively compensating each other in our device measurements, the most likely scenario is that loss is dominated by mechanisms outside the metal-air top surface and metal-substrate interface. These mechanisms should still pertain to material interfaces based on the increase in $\delta_{tls}$ between $w = 30\ \mu m$ and $w = 3\ \mu m$ with the participation ratio of the main interfaces. One of the two possible candidates is the metal-air interface along the sidewalls, which likely hosts additional microscopic structural and chemical defects induced by the dry-etch process [28, 29]. If the loss tangent was dominated by the contribution from these defects, the resonator quality factor would become insensitive to the thickness and composition of the metal oxide layer. This would be consistent with the recent observation that suppressing the top native oxide layer of Ta resonators using gold capping does not result in an improvement of the quality factor [30]. The other candidate is the substrate-air interface. In addition to dry-etch induced damages, we also expect several angstroms of silicon oxide to have regrown in the day between the end-of-line buffered oxide etch and device loading in the dilution refrigerator [31]. Given the higher reported values for the loss tangent of silicon oxide [31] than tantalum oxide [16], it is plausible that the silicon oxide layer dominates dielectric loss for both devices, regardless of the sputtering method used for Ta.

While we cannot provide a definite explanation for the comparable performance of the two types of films based on our current data, the fact that there is no clear correlation between a number of material properties and resonator performance helps narrow down the range of candidate mechanisms for dominant sources of loss in state-of-the-art Ta superconducting resonators. It is corroborated by other recent studies that have demonstrated similar resonator performance for different grain sizes of Ta on sapphire [32], and quality factors in the $10^6$ range for polycrystalline Ta grown on silicon at cryogenic temperatures [33]. Our results not only highlight the robustness of the low loss of $\alpha$-Ta with regards to the film structural and transport properties, but they also suggest that the thickness and disorder of the surface metal oxides do not correlate with device performance [25, 34] for Ta-based devices in the low $10^{-7}$ single-photon-level loss regime.

## IV. CONCLUSION

We have shown that resonators fabricated from $\alpha$-tantalum films grown on silicon at room temperature on a Nb seed layer can perform as well as the current state-of-the-art resonators fabricated from films grown at high temperature. Irrespective of the growth process temperature, our resonators yield state-of-the-art losses in the single photon regime, with median values in the low $10^{-7}$. Profound differences are found in the material properties of these two types of films, including a three-orders-of-magnitude difference in their grain size, significant oxygen diffusion from the metal-air interface in the room-temperature film, a one-order-of-magnitude difference in RRR and kinetic inductance, and interdiffusion at the metal-substrate interface in the high-temperature film. While we cannot rule out that the impact from these differences in material properties on TLS loss may compensate each other, the more likely scenario is that TLS loss in our Ta-based superconducting circuits is dominated either by the fabrication-induced defects along the sidewalls, and/or by the surface oxides and fabrication-induced defects at the substrate-air interface. In addition to highlighting potential areas of further improvements for Ta-based superconducting circuits, our study shows that $\alpha$-Ta sputtered at room temperature is a potent material for integration into high-performance superconducting qubits. Its use should be particularly beneficial for applications that require either a high process throughput or compatibility with low-thermal budget fabrication processes, and which can accommodate the larger kinetic inductance.

## V. METHODS

### A. Fabrication

The films were deposited using magnetron DC sputtering on 4-inch high-resistivity silicon wafers (>20 k$\Omega$cm). Prior to deposition, the wafers were treated in a buffered oxide etchant (BOE 10:1) solution to remove the native oxide layer. The base pressure of the deposition chamber was under $2\times10^{-7}$ Pa, and the purity of the tantalum and niobium targets were respectively 99.998% and 99.99%. The substrate was either heated at 350 °C (as measured at the substrate) or kept at room temperature during the deposition of the HT and RT films, respectively. The total thickness of the films is 100 nm, which includes a 5-nm thick Nb buffer layer for the RT films.

The metal-coated wafers were then cleaned with an oxygen plasma for 10 minutes at 150W, dehydrated on a hotplate at 140 °C, and coated with Vapor Hexamethyldisilazane (HMDS) before being baked again at 140 °C. The SPR955 resist was spun and baked at 100°C, patterned with a direct write optical lithography method (MLA150 Heidelberg), and developed with a MF319 solution. The films were finally etched with a two-step sequence including Ar milling and inductively coupled plasma reactive ion etching (ICP-RIE) etch with chlorine-based chemistry, and cleaned with an oxygen plasma followed by an EKC chemical treatment. After dicing the wafers, the chips selected for cryogenic RF measurements were again treated in BOE 10:1 prior to packaging.

### B. Film characterization

The structural properties of the films were first characterized with x-ray diffraction (XRD) using Cu K$\alpha$ radiation in parallel-beam configuration on a Rigaku Smartlab diffractometer. The morphologies and elemental distributions at the interfaces of resonator devices were imaged using scanning transmission electron microscopy and electron energy loss spectroscopy (STEM-EELS) with an FEI probe-corrected TITAN operated at 200 kV, on 70-nm thick lamellae prepared with the plasma focused ion beam (P-FIB) lift-out technique. Depth profiles of the Ta oxide species near the film surfaces were obtained using variable kinetic energy x-ray photoelectron spectroscopy (VKE-XPS). The VKE-XPS measurements were performed at the NIST Spectroscopy Soft and Tender (SST-1 and SST-2) beamlines at the National Synchrotron Light Source II in Brookhaven National Laboratory. The grains in the films were characterized with a Dimension Icon atomic force microscope (AFM) from Bruker in a PeakForce scanning mode, and through electron backscattering diffraction (EBSD) and transmission Kikuchi diffraction (TKD) in a Thermo Fisher Helios 5 UC scanning electron microscope (SEM) equipped with an e-flash FS detector and EBSD or Optimus 2 TKD detector head. The transport measurements were performed in a Quantum Design DynaCool physical property measurement system (PPMS), from room-temperature to 1.8 K, and with a magnetic field oriented along the normal of the samples.

### C. Device design and measurements

The 10 mm x 10 mm resonator chips consist of twelve side-coupled $\lambda/4$ CPW resonators connected to one transmission line. The measurements were carried out using a vector network analyzer (VNA) in a conventional setup [35], which consists of 75 dB total input attenuation on the input line and a cryogenic low noise amplifier followed by a room-temperature amplifier on the output line. The internal quality factors ($Q_i$) or loss ($\delta_i$) were extracted from the measurements of the transmission signal $S_{21}$ of transmission lines inductively coupled to resonators and analyzed with a diameter correction method (DCM) method described in Ref. [36]. The losses presented in this study result from 15 to 70 repeated measurements of each resonator on the chips, for both materials and designs. The loss measured in the 0.5 to 2 photon number range are reported as single photon in the main text.

ACKNOWLEDGMENTS

This research used the NIST-operated Spectroscopy Soft and Tender Beamlines (SST-1 and SST-2) of the National Synchrotron Light Source II, U.S. Department of Energy (DOE) Office of Science User Facilities operated for the DOE Office of Science by Brookhaven National Laboratory under Contract No. DE-SC0012704. Certain commercial equipment, instruments, or materials are identified in this paper in order to specify the experimental procedure adequately, and do not represent an endorsement by the National Institute of Standards and Technology. We thank Simone Severini, Bill Vass, James Hamilton, Peter DeSantis, Fernando Brandao, Eric Chisholm, Matthew Matheny, and Shmulik Eisenmann at AWS for their involvement and support of the research activities at the AWS Center for Quantum Computing.

[1] J. G. C. Milne, "Superconducting Transition Temperature of High-Purity Tantalum Metal," Physical Review, vol. 122, pp. 387–388, Apr. 1961.

[2] A. P. M. Place, L. V. H. Rodgers, P. Mundada, B. M. Smitham, M. Fitzpatrick, Z. Leng, A. Premkumar, J. Bryon, A. Vrajitoarea, S. Sussman, G. Cheng, T. Madhavan, H. K. Babla, X. H. Le, Y. Gang, B. Jäck, A. Gyenis, N. Yao, R. J. Cava, N. P. de Leon, and A. A. Houck, "New material platform for superconducting transmon qubits with coherence times exceeding 0.3 milliseconds," Nature Communications, vol. 12, p. 1779, Dec. 2021.

[3] C. Wang, X. Li, H. Xu, Z. Li, J. Wang, Z. Yang, Z. Mi, X. Liang, T. Su, C. Yang, G. Wang, W. Wang, Y. Li, M. Chen, C. Li, K. Linghu, J. Han, Y. Zhang, Y. Feng, Y. Song, T. Ma, J. Zhang, R. Wang, P. Zhao, W. Liu, G. Xue, Y. Jin, and H. Yu, "Towards practical quantum computers: transmon qubit with a lifetime approaching 0.5 milliseconds," npj Quantum Information, vol. 8, pp. 1–6, Jan. 2022. Number: 1 Publisher: Nature Publishing Group.

[4] S. Ganjam, Y. Wang, Y. Lu, A. Banerjee, C. U. Lei, L. Krayzman, K. Kisslinger, C. Zhou, R. Li, Y. Jia, M. Liu, L. Frunzio, and R. J. Schoelkopf, "Surpassing millisecond coherence in on chip superconducting quantum memories by optimizing materials and circuit design," Nature Communications, vol. 15, p. 3687, May 2024. Publisher: Nature Publishing Group.

[5] R. A. McLellan, A. Dutta, C. Zhou, Y. Jia, C. Weiland, X. Gui, A. P. M. Place, K. D. Crowley, X. H. Le, T. Madhavan, Y. Gang, L. Baker, A. R. Head, I. Waluyo, R. Li, K. Kisslinger, A. Hunt, I. Jarrige, S. A. Lyon, A. M. Barbour, R. J. Cava, A. A. Houck, S. L. Hulbert, M. Liu, A. L. Walter, and N. P. de Leon, "Chemical Profiles of the Oxides on Tantalum in State of the Art Superconducting Circuits," Advanced Science, vol. 10, no. 21, p. 2300921, 2023.

[6] D. W. Face, S. T. Ruggiero, and D. E. Prober, "Ion-beam deposition of Nb and Ta refractory superconducting films," Journal of Vacuum Science & Technology A: Vacuum, Surfaces, and Films, vol. 1, pp. 326–330, Apr. 1983.

[7] D. W. Face and D. E. Prober, "Nucleation of body-centered-cubic tantalum films with a thin niobium underlayer," Journal of Vacuum Science & Technology A: Vacuum, Surfaces, and Films, vol. 5, pp. 3408–3411, Nov. 1987.

[8] L. D. Alegria, D. M. Tennant, K. R. Chaves, J. R. I. Lee, S. R. O'Kelley, Y. J. Rosen, and J. L. DuBois, "Two-level systems in nucleated and non-nucleated epitaxial alpha-tantalum films," Applied Physics Letters, vol. 123, p. 062601, Aug. 2023.

[9] M. Singer, B. Schoof, H. Gupta, D. Zahn, J. Weber, and M. Tornow, "Tantalum thin films sputtered on silicon and on different seed layers: material characterization and coplanar waveguide resonator performance," Sept. 2024. arXiv:2409.06041 [cond-mat].

[10] R. Barends, J. Baselmans, J. Hovenier, J. Gao, S. Yates, T. Klapwijk, and H. Hoevers, "Niobium and Tantalum High Q Resonators for Photon Detectors," IEEE Transactions on Applied Superconductivity, vol. 17, pp. 263–266, June 2007.

[11] Y. Urade, K. Yakushiji, M. Tsujimoto, T. Yamada, K. Makise, W. Mizubayashi, and K. Inomata, "Microwave characterization of tantalum superconducting resonators on silicon substrate with niobium buffer layer," APL Materials, vol. 12, p. 021132, Feb. 2024.

[12] S. Poorgholam-Khanjari, V. Seferai, P. Foshat, C. Rose, H. Feng, R. H. Hadfield, M. Weides, and K. Delfanazari, "Engineering high-Q superconducting tantalum microwave coplanar waveguide resonators for compact coherent quantum circuits," Dec. 2024. arXiv:2412.16099 [quant-ph].

[13] D. P. Lozano, M. Mongillo, X. Piao, S. Couet, D. Wan, Y. Canvel, A. M. Vadiraj, T. Ivanov, J. Verjauw, R. Acharya, J. Van Damme, F. A. Mohiyaddin, J. Jussot, P. P. Gowda, A. Pacco, B. Raes, J. Van De Vondel, I. P. Radu, B. Govoreanu, J. Swerts, A. Potočnik, and K. De Greve, "Low-loss -tantalum coplanar waveguide resonators on silicon wafers: fabrication, characterization and surface modification," Materials for Quantum Technology, vol. 4, p. 025801, June 2024.

[14] L. Shi, T. Guo, R. Su, T. Chi, Y. Sheng, J. Jiang, C. Cao, J. Wu, X. Tu, G. Sun, J. Chen, and P. Wu, "Tantalum microwave resonators with ultra-high intrinsic quality factors," Applied Physics Letters, vol. 121, p. 242601, Dec. 2022.

[15] K. Grigoras, N. Yurttagül, J.-P. Kaikkonen, E. T. Mannila, P. Eskelinen, D. P. Lozano, H.-X. Li, M. Rommel, D. Shiri, N. Tiencken, S. Simbierowicz, A. Ronzani, J. Hätinen, D. Datta, V. Vesterinen, L. Gron-

berg, J. Biznarova, A. F. Roudsari, S. Kosen, A. Osman, M. Prunnila, J. Hassel, J. Bylander, and J. Govenius, "Qubit-Compatible Substrates With Superconducting Through-Silicon Vias," IEEE Transactions on Quantum Engineering, vol. 3, pp. 1–10, 2022.

[16] K. D. Crowley, R. A. McLellan, A. Dutta, N. Shumiya, A. P. Place, X. H. Le, Y. Gang, T. Madhavan, M. P. Bland, R. Chang, N. Khedkar, Y. C. Feng, E. A. Umbarkar, X. Gui, L. V. Rodgers, Y. Jia, M. M. Feldman, S. A. Lyon, M. Liu, R. J. Cava, A. A. Houck, and N. P. de Leon, "Disentangling Losses in Tantalum Superconducting Circuits," Physical Review X, vol. 13, p. 041005, Oct. 2023. Publisher: American Physical Society.

[17] W. A. Phillips, "Two-level states in glasses," Reports on Progress in Physics, vol. 50, pp. 1657–1708, Dec. 1987.

[18] P. w. Anderson, B. I. Halperin, and c. M. Varma, "Anomalous low-temperature thermal properties of glasses and spin glasses," The Philosophical Magazine: A Journal of Theoretical Experimental and Applied Physics, vol. 25, pp. 1–9, Jan. 1972.

[19] J. Burnett, L. Faoro, and T. Lindström, "Analysis of high quality superconducting resonators: consequences for TLS properties in amorphous oxides," Superconductor Science and Technology, vol. 29, p. 044008, Apr. 2016.

[20] R. Gao, W. Yu, H. Deng, H.-S. Ku, Z. Li, M. Wang, X. Miao, Y. Lin, and C. Deng, "Epitaxial titanium nitride microwave resonators: Structural, chemical, electrical, and microwave properties," Physical Review Materials, vol. 6, p. 036202, Mar. 2022. Publisher: American Physical Society.

[21] C. E. Murray, "Material matters in superconducting qubits," Materials Science and Engineering: R: Reports, vol. 146, p. 100646, Oct. 2021.

[22] A. J. Annunziata, D. F. Santavicca, L. Frunzio, G. Catelani, M. J. Rooks, A. Frydman, and D. E. Prober, "Tunable superconducting nanoinductors," Nanotechnology, vol. 21, p. 445202, Nov. 2010.

[23] A. Glezer Moshe, E. Farber, and G. Deutscher, "Granular superconductors for high kinetic inductance and low loss quantum devices," Applied Physics Letters, vol. 117, Aug. 2020. Publisher: AIP Publishing.

[24] J. Mun, P. V. Sushko, E. Brass, C. Zhou, K. Kisslinger, X. Qu, M. Liu, and Y. Zhu, "Probing Oxidation-Driven Amorphized Surfaces in a Ta(110) Film for Superconducting Qubit," ACS Nano, vol. 18, pp. 1126–1136, Jan. 2024. Publisher: American Chemical Society.

[25] A. Premkumar, C. Weiland, S. Hwang, B. Jäck, A. P. M. Place, I. Waluyo, A. Hunt, V. Bisogni, J. Pelliciari, A. Barbour, M. S. Miller, P. Russo, F. Camino, K. Kisslinger, X. Tong, M. S. Hybertsen, A. A. Houck, and I. Jarrige, "Microscopic relaxation channels in materials for superconducting qubits," Communications Materials, vol. 2, pp. 1–9, July 2021. Publisher: Nature Publishing Group.

[26] T. Nakayama and K. Kobinata, "Physics of Schottky-barrier change by segregation and structural disorder at metal/Si interfaces: First-principles study," Thin Solid Films, vol. 520, pp. 3374–3378, Feb. 2012.

[27] J. Y. Cheng, M. H. Wang, and L. J. Chen, "Formation of Amorphous Interlayers by Solid-State Diffusion in Refractory Metal/Silicon Systems," MRS Online Proceedings Library, vol. 187, pp. 77–82, Dec. 1990.

[28] K. Uejima and T. Umeda, "Microscopic origins of dry-etching damages in silicon large-scaled integrated circuits revealed by electrically detected magnetic resonance," Applied Physics Letters, vol. 104, p. 082111, Feb. 2014.

[29] Y. Cao, G. Pu, H. Cao, R. Zhan, F. Jin, M. Dan, Z. Xu, K. Zhang, J. Nie, and Y. Wang, "Surface quality and microstructure evolution in fused silica under SF6/Ar reactive ion beam etching," Journal of Non-Crystalline Solids, vol. 641, p. 123144, Oct. 2024.

[30] R. D. Chang, N. Shumiya, R. A. McLellan, Y. Zhang, M. P. Bland, F. Bahrami, J. Mun, C. Zhou, K. Kisslinger, G. Cheng, A. C. Pakpour-Tabrizi, N. Yao, Y. Zhu, M. Liu, R. J. Cava, S. Gopalakrishnan, A. A. Houck, and N. P. de Leon, "Eliminating Surface Oxides of Superconducting Circuits with Noble Metal Encapsulation," Aug. 2024. arXiv:2408.13051 [cond-mat, physics:quant-ph].

[31] M. Morita, T. Ohmi, E. Hasegawa, M. Kawakami, and M. Ohwada, "Growth of native oxide on a silicon surface," Journal of Applied Physics, vol. 68, pp. 1272–1281, Aug. 1990.

[32] S. G. Jones, N. Materise, K. W. Leung, B. D. Isakov, X. Chen, J. Zheng, A. Gyenis, B. Jaeck, and C. R. H. McRae, "Grain size in low loss superconducting Ta thin films on c-axis sapphire," July 2023. arXiv.2307.11667.

[33] T. A. J. van Schijndel, A. P. McFadden, A. N. Engel, J. T. Dong, W. J. Yánez-Parreño, M. Parthasarathy, R. W. Simmonds, and C. J. Palmstrøm, "Cryogenic growth of tantalum thin films for low-loss superconducting circuits," May 2024. arXiv:2405.12417 [cond-mat].

[34] M. V. P. Altoé, A. Banerjee, C. Berk, A. Hajr, A. Schwartzberg, C. Song, M. Alghadeer, S. Aloni, M. J. Elowson, J. M. Kreikebaum, E. K. Wong, S. M. Griffin, S. Rao, A. Weber-Bargioni, A. M. Minor, D. I. Santiago, S. Cabrini, I. Siddiqi, and D. F. Ogletree, "Localization and Mitigation of Loss in Niobium Superconducting Circuits," PRX Quantum, vol. 3, p. 020312, Apr. 2022.

[35] C. R. H. McRae, H. Wang, J. Gao, M. R. Vissers, T. Brecht, A. Dunsworth, D. P. Pappas, and J. Mutus, "Materials loss measurements using superconducting microwave resonators," Review of Scientific Instruments, vol. 91, p. 091101, Sept. 2020.

[36] M. S. Khalil, M. J. A. Stoutimore, F. C. Wellstood, and K. D. Osborn, "An analysis method for asymmetric resonator transmission applied to superconducting devices," Journal of Applied Physics, vol. 111, p. 054510, Mar. 2012. arXiv:1108.3117 [cond-mat].